\documentclass[fleqn,12pt,epsfig,twoside]{article}
\usepackage{espcrc1}

\usepackage{epsfig}
\usepackage[figuresright]{rotating}

\newcommand{\AmS}{{\protect\the\textfont2
  A\kern-.1667em\lower.5ex\hbox{M}\kern-.125emS}}

\title{ Dilepton and vector meson production in elementary
and in heavy ion reactions} 

\author{C. Fuchs\address[tue]{Institut f\"ur Theoretische Physik,
Universit\"at T\"ubingen, D-72076 T\"ubingen, Germany }, 
Amand Faessler\addressmark[tue], 
D. Cozma\addressmark[tue], B.V. Martemyanov\addressmark[tue]
\address[mos]{Institute for Theoretical and Experimental Physics, 
117259 Moscow, Russia}, 
M. Krivoruchenko\addressmark[tue]\addressmark[mos]
}

\begin{document}

\maketitle

\begin{abstract}
We present a unified description of the vector 
meson and dilepton production in elementary 
and in heavy ion reactions. The production of vector mesons 
is described via the excitation of nucleon 
resonances. Medium effects in heavy ion reactions are discussed.
\end{abstract}

\vspace*{0.5cm}
Dilepton spectra from heavy-ion collisions are considered 
as a suitable tool to study medium modifications of vector 
mesons ($\rho,\omega,\phi$) in a dense nuclear environment 
which is produced in heavy ion reactions. 
Suchs medium effects manifest themselves in the modification of 
widths and masses of resonances produced in nuclear collisions.
E.g., the Brown-Rho scaling \cite{BR} is equivalent to a reduction 
of the $\rho$ meson masses in the nuclear medium. The same conclusion 
is obtained from QCD sum rules \cite{QCDSR}. 
Hadronic models \cite{KKW,rapp} based 
on dispersion analyses of forward scattering amplitudes predict  
Vector meson mass shifts are in 
general small and positive, whereas at low momenta they can change 
the sign which is in qualitative agreement with the 
Brown-Rho scaling and the results from QCD sum rules. 
Indeed, at CERN a significant enhancement of the low-energy 
dilepton yield below the 
$\rho $ and $\omega $ peaks \cite{ceres} in heavy reaction systems 
compared to light systems and proton induced reactions 
has been observed  \cite{ceres}. 
Theoretically, this enhancement 
can be explained within a hadronic picture by the assumption of 
a dropping $\rho $ mass or, alternatively, by the 
formation of a quark-gluon plasma \cite{rapp}. 
A similar situation occurs at a completely different energy scale,
 namely around 1 A.GeV incident energies where the low mass region
of dilepton spectra measured by the 
DLS Collaboration at the BEVALAC \cite{DLS}
are underestimated by present transport
calculations compared to $pp$ and $pd$ reactions 
\cite{ernst,BCRW98,shekther03}. 
However, in contrast to ultra-relativistic reactions (SPS) 
the situation does not improve when full spectral functions
and/or a dropping mass of the vector mesons are taken
into account. This fact is known as 
the DLS {\it puzzle}. 
\begin{figure}[h]
\begin{minipage}[h]{95mm}
\unitlength1cm
\begin{picture}(9.,3.5)
\put(0.5,0){\makebox{\epsfig{file=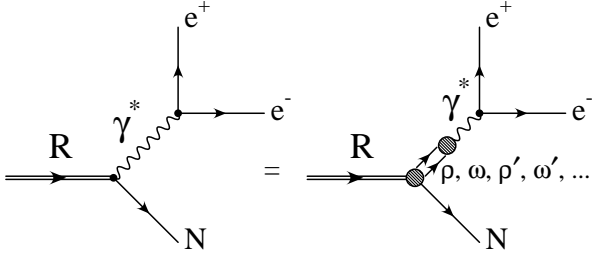,width=9.0cm}}}
\end{picture}
\end{minipage}
\vspace*{-1.5cm}
\hspace*{0.0cm}
\begin{minipage}[h]{60mm}
\caption{Decay of nuclear resonances to dileptons in the extended VMD 
model. The $RN\gamma$ transition form factors contain contributions from 
ground state and excited $\rho$ and $\omega$ mesons.
}
\end{minipage}
\label{graph1_fig}
\vspace*{1.5cm}
\end{figure}

For all these studies a precise and rather complete knowledge of the relative
weights for existing decay channels is indispensable in order to
draw reliable conclusions from dilepton spectra. 
we develloped a unified model for the description of vector 
mesons and dilepton pairs in elmentary nucleon-nucleon and 
pion-nucleon reactions and in heavy ino reactions. The model 
is based on an extension of the vector meson dominance (eVMD) model 
which describes  meson decay channels \cite{krivo00}, 
including channels which have been neglected so far, 
such as e.g. four-body decays $\rho^0\rightarrow\pi^0\pi^0e^+e^-$. 
In \cite{krivo02} a fully relativistic and kinematically complete 
treatment of vector meson $R\rightarrow N\rho(\omega)$ and 
dilepton decays $R\rightarrow N\ e^+\ e^-$ of nucleon 
resonances with arbitrary spin and parity was performed. The magnetic, 
electric, and Coulomb transition form factors have been determined 
 fitting available photo- and electro-production data. 
The resonance model schematically depicted in Fig. 1 provides an accurate description of 
exclusive vector meson production in nucleon-nucleon 
collisions $NN\rightarrow NN\rho(\omega)$ as well as in pion scattering 
$\pi N\rightarrow N\rho(\omega)$ \cite{shekther03,omega02} and has been 
succesfully applied to $\phi$ production \cite{phi03} as well as to the 
dilepton production in $pp$ reactions at BEVALAC 
energies  \cite{resdec}. 

Fig. 2  shows the 
$\omega$ production in elementary $NN$ reactions. The different cross 
sections are shown as functions of the excess energy $\epsilon$. 
As discussed in \cite{omega02}, the 
resonance model (with a large $N^*(1535)N\omega$ coupling) leads to 
very accurate description of the measured on-shell cross section. 
It has, however, a very strong off-shell component which fully 
contributes to the dilepton production. The weak coupling scenario, on 
the other side, has only small off-shell component but the 
reproduction of the data is relatively poor in the low energy regime. 
\begin{figure}[h]
\begin{minipage}[h]{95mm}
\unitlength1cm
\begin{picture}(9.,7.5)
\put(0.5,0){\makebox{\epsfig{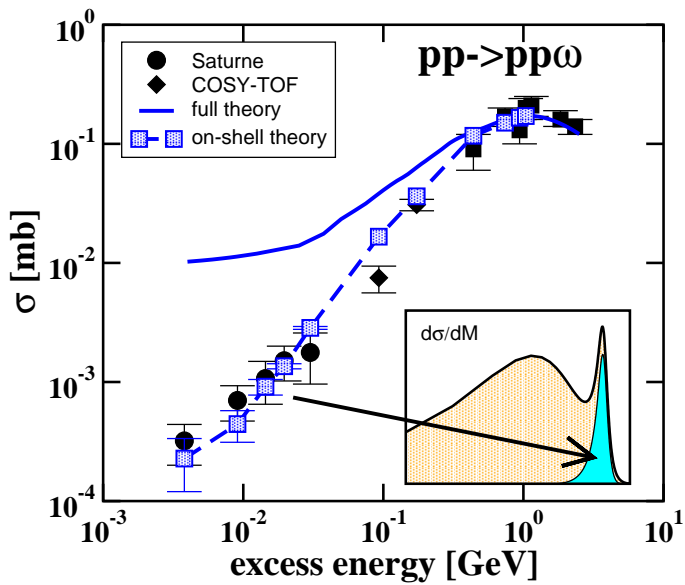}}}
\end{picture}
\end{minipage}
\vspace*{-1.5cm}
\hspace*{0.5cm}
\begin{minipage}[h]{55mm}
\caption{Exclusive $pp\rightarrow pp\omega$ cross section 
obtained in the resonance model as a function of the excess energy 
$\epsilon$. The solid curve shows the full cross 
section including off-shell 
contributions while the squares show the experimentally 
detectable on-shell part of the cross section. 
Data are taken from \protect\cite{omdat}. 
}
\end{minipage}
\vspace*{1.0cm}
\label{sigom_fig}
\end{figure}

In \cite{omega04} it has finally been demonstrated that the 
resonance model is able to describe the measured $\omega$- and $\phi$-meson 
angular distributions in proton-proton reactions. The assumption of 
dominant contributions from the 
$N^*(1720)\frac{3}{2}^+$ and $N^*(1900)\frac{3}{2}^+$
resonances at $\sqrt{s} = 2.83$ GeV where data from COSY-TOF have been taken 
yields the right pattern for the $\omega$ angular distribution as can be 
seen from Fig. 3.
\begin{figure}[h]
\begin{minipage}[h]{85mm}
\unitlength1cm
\begin{picture}(8.,7.0)
\put(0.5,0){\makebox{\epsfig{file=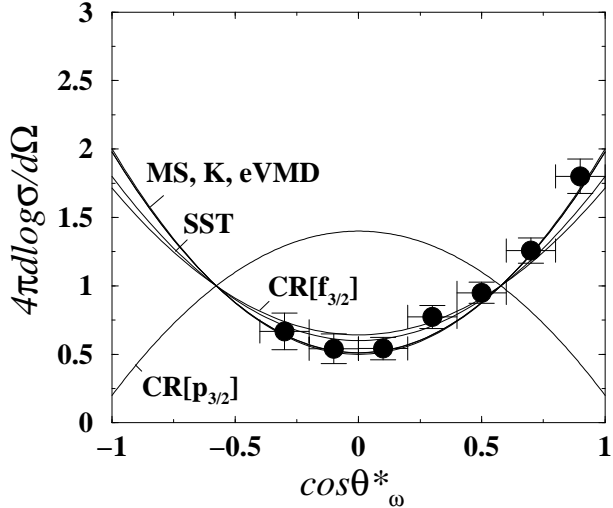,width=8.0cm}}}
\end{picture}
\end{minipage}
\vspace*{-1.5cm}
\hspace*{0.5cm}
\begin{minipage}[h]{65mm}
\caption{$\omega$-meson angular distribution in $pp$ reactions 
at an excess energy $\epsilon =173$ MeV assuming that 
the reaction goes through the $N^*(1900)\frac{3}{2}^+$ 
resonance. The experimental data are from COSY-TOF \cite{omdat}. 
Results from various partial wave analyses and quark models are compared 
to the present eVMD model.
}
\end{minipage}
\vspace*{1.0cm}
\label{sigom_fig}
\end{figure}

Although the present model is able to reproduce the vector meson and the 
dilepton production in elementary reactions with high precision the 
situation in unsatisfactory when turning to heavy ion collisions. 
Heavy ion collisions are described 
within the QMD transport model \cite{shekther03}. 
Without additional in-medium effect  we observe in two distinct 
kinematical regions significant deviations from 
the dilepton yields measured by the DLS Collaboration in 
$C+C$ and $Ca+Ca$ reactions at 1 AGeV. As can be seen from Fig. 4 at small 
invariant masses the experimental data 
are strongly underestimated which confirms the observations made 
by other groups \cite{ernst,BCRW98}. Although accounting for the experimental resolution 
we observe further a clear structure of the $\rho/\omega$ peak which is  
not present in the data. 
\begin{figure}[h]
\unitlength1cm
\begin{picture}(13.,7.0)
\put(0.5,0){\makebox{\epsfig{file=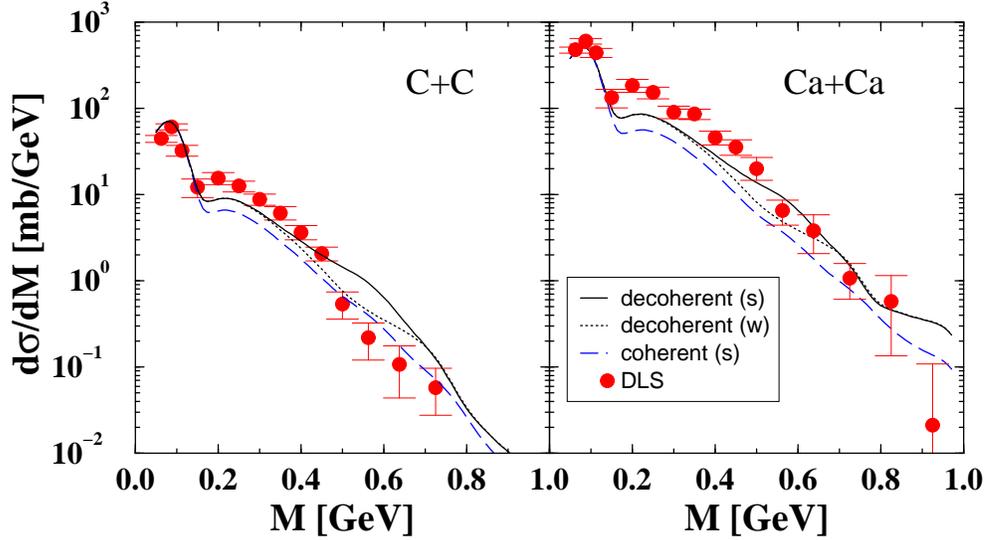,width=13.0cm}}}
\end{picture}
\vspace*{-1.0cm}
\caption{Influence of the microscopically determined decoherent dilepton emission 
in $C+C$ and $Ca+Ca$ reactions. A strong (s), respectively, weak (w) 
$N^*(1535)-N\omega $ coupling is used. For comparison also the 
coherent case (s) is shown. Data are from the DLS Collaboration \protect\cite{DLS}.
}
\label{DLS_AA_dec2}
\end{figure}
The collisional broadening of the vector mesons suppresses the  
$\rho/\omega$ peak in the dilepton spectra. This allows to extract  
empirical values for the in-medium widths of the vector mesons. From  
the reproduction of the DLS data the following estimates for the 
collision widths $\Gamma_{\rho}^{\rm coll} = 150$ MeV and 
$\Gamma_{\omega}^{\rm coll} = 100 - 300$ MeV can be made. The in-medium values 
correspond to an average nuclear density of about 1.5 $\rho_0$ and have been
used in the calculation shown in Fig. 4. The 
forthcoming data from HADES \cite{hades} will 
certainly help to constrain these values with higher precision. 

The second medium effect discussed in \cite{shekther03} concerns the problem of 
quantum interference. Semi-classical transport models like QMD do generally 
not account for interference effects, i.e. they propagate probabilities 
rather than amplitudes and assume that relative phases cancel the 
interference on average. 
However, interference effects can play an important role for the dilepton 
production. In the present model the decay of nuclear resonances which is 
the dominant source for the dilepton yield, requires the destructive interference 
of intermediate $\rho$ and $\omega$ mesons with their excited states. 
The interference can at least partially be destroyed by the presence of 
the medium which leads to an enhancement of the corresponding dilepton 
yield (see Fig. 4). We proposed a scheme to treat the decoherence in the medium on 
a microscopic level. The account for decoherence improves the agreement with 
the DLS data in the low mass region. However, to attack the fundamental 
question of medium modifications of vector mesons seriously 
and to come to firm conclusions needs, besides 
further theoretical efforts, much more high precision data. This will 
be the task for HADES in the next years.

\end{document}